\documentstyle[amssymb,aps,prl,twocolumn]{revtex}

\input{tcilatex}

\begin{document}
\title{Positive magnetoresistance and orbital ordering in La$_{1-x}$Sr$_{x}$MnO$%
_{3} $}
\author{M. Paraskevopoulos$^1$, J. Hemberger$^1$, A. Loidl$^1$, A. A. Mukhin$^2$, V.
Yu. Ivanov$^2$ and A. M. Balbashov$^3$}
\address{$^{1}$Experimentalphysik V, Elektronische Korrelationen und Magnetismus, \\
Institut f\"{u}r Physik, Universit\"{a}t Augsburg, D - 86159 Augsburg,\\
Germany\\
$^{2}$General Physics Institute of the Russian Acad. Sci., 117942 Moscow,\\
Russia\\
$^3$Moscow Power Engineering Institute, 105835 Moscow, Russia}
\date{submitted to PRL \today{}}
\maketitle

\begin{abstract}
We report on detailed transprort measurements of single crystalline La$%
_{1-x} $Sr$_{x}$MnO$_{3}$ ($x\leqslant 0.2$). We have found a giant positive
magnetoresistance in the compositions range between $0.1\leqslant x\leqslant
0.125$ and give an explaination in terms of orbital ordering due to the
interplay between superexchange interactions and Jahn-Teller distortions.
\end{abstract}



\vspace{4ex}





In the last years an overwhelming interest in the manganite perovskites La$%
_{1-x}$A$_{x}$MnO$_{3}$ arose primarily due to the observation of a collosal
negative magnetoresistance (CMR) close to $x=0.3$ \cite{Chahara}. These CMR
effects at the ferromagnetic (FM) phase transition were explained within
extended double-exchange (DE) models \cite{Zener,deGennes,Furukawa}. Since
then it became clear that this compounds reveal many interesting and
puzzling phenomena which can not be accounted for by double exchange alone.
It was Millis et al. \cite{Millis} who first related dynamic Jahn-Teller
(JT) distortions to the CMR effect. In a recent paper \cite{Paras} we have
presented a detailed phase diagram of La$_{1-x}$Sr$_{x}$MnO$_{3}$ at low Sr
concentrations. In the concentration regime $0.1\leqslant x\leqslant 0.15$
we detected a ferromagnetic (FM) and insulating (I) ground state which is
followed by a canted antiferromagnetic (CA) or mixed phase at elevated
temperatures \cite{Paras}. This insulating FM-phase results from
superexchange (SE) interactions in a charge-ordered (CO)\ phase which
probably also reveals orbital order.

In this paper we will present detailed transport measurements on La$_{1-x}$Sr%
$_{x}$MnO$_{3}$ single crystals. It will be shown that the double degeneracy
of the $e_{g}$ orbitals and their ordering in real space must be considered
in order to understand the ferromagnetic and insulating ground state in this
doping regime. Recently theoretical models considered the orbital degrees of
freedom, but have mainly focused on the properties of the LaMnO$_{3}$ and La$%
_{0.5}$Sr$_{0.5}$MnO$_{3}$ compounds \cite{Koshibae}. Although Ahn and
Millis \cite{Ahn} discussed the interplay of orbital and charge ordering for
La$_{0.875}$Sr$_{0.125}$MnO$_{3}$ they did not considered the effect of this
types of ordering on the mangetic properties. Using a mean field
approximation Maezono $et$ $al.$ \cite{Maezono}\ have obtained an overall
phase diagram of doped manganites based on the interplay of SE- and
DE-interactions.

Here we provide strong experimental evidence that the competition between
JT-, SE- and DE-interactions is responsible for the rich variety of magnetic
and structural phase transitions in the low doping phase diagram ($%
x\leqslant 0.2$). At the phase boundary of the insulating FM ground state, a
strong (collossal) positive MR appears. We will show that the close coupling
of structural and magnetic phase transitions and especially the magnetic
field induced structural transitions can only be explained assuming orbital
order in the O%
\'{}%
\'{}%
-phase.

Single crystals of La$_{1-x}$Sr$_{x}$MnO$_{3}$, with concentrations $%
0.1\leqslant x\leqslant 0.2$ were grown by a floating zone method with
radiation heating in air atmosphere. X-ray diffraction of crushed single
crystals revealed high-quality single-phase materials. X-ray topography
indicated twinning of the crystals. Transport measurements in the
temperature range 1.5-300 K were performed with the standard four-probe
method in fields up to 14 T.

The relevant part of the phase diagram around $x=0.125$ is reproduced in
Fig. 1. Close to this concentration a rather unusual sequence of phase
transitions has been detected. At room temperature La$_{0.875}$Sr$_{0.125}$%
MnO$_{3}$ is orthorhombic. A long range JT distortion (and hence orbital
order) appears at T$_{\text{OO%
\'{}%
}}$=265 K, both orthorhombic phases beeing insulating \cite{Mukhin}. At T$%
_{CA}$=180 K a CA structure is established where the resistivity is reduced
almost by a factor of 10. It is this regime which has been treated as a FM
and metallic phase during the last years. This however was not based on
experimental results but only due to the fact that $d\rho /dT>0$ in a
limited temperature regime and that a FM regime which is followed by a
canted spin structure at lower temperatures nicely fits into de Gennes phase
diagram \cite{deGennes}. We have shown that this phase reveals a canted
structure \cite{Paras}. Of course, we also can not exclude a mixed phase in
this regime, but it is definitely not metallic. The decrease in the
resistivity probably indicates an increase of the charge transfer matrix
elements due to the increasing importance of the DE mechanism combined with
the freezing out of spin-disorder. Finally at T$_{\text{O%
\'{}%
O%
\'{}%
\'{}%
}}$=150 K and T$_{C}$=140 K a structural phase transition is immediately
followed by the onset of FM order\cite{Paras}. The O%
\'{}%
\'{}%
-phase reveals charge order in an almost pseudocubic lattice and the ground
state is a FM insulator.

The field dependence of the magnetoresistance ($\Delta \rho $/$\rho $(0)=$%
\rho $(H)-$\rho $(0)/$\rho $(0)) for $x=0.1$ and $x=0.125$ is shown in Fig.
2 for various temperatures. For temperatures T%
\mbox{$>$}%
T$_{\text{O%
\'{}%
O%
\'{}%
\'{}%
}}$ we find negative MR effects. However we would like to point out that
here the negative MR appears at the transition from a paramagnetic (PM)
insulator into a CA or mixed phase. Small negative MR effects again appear
for T%
\mbox{$<$}%
T$_{C}$=105 K. However, between T$_{\text{O%
\'{}%
O%
\'{}%
\'{}%
}}$ and T$_{C}$ large or even colossal positive MR effects appear, due to
the fact that the FM ground state is indeed more insulating than the CA
phase. A closer inspection of Fig. 2 reveals that two subsequent jumps
appear as a function of increasing field. Both jumps induce higher
resistivity states and correspond to those ones observed in the
magnetization curves (see Fig. 3 in \cite{Paras}). The first jump is due to
a field induced structural phase transition from the JT-distorted O%
\'{}%
- to the pseudocubic O%
\'{}%
\'{}%
- orthorhombic phase and obviously is accompanied by a real space charge
ordering of the Mn$^{3+}$ and Mn$^{4+}$ ions \cite{Yam97}. At the subsequent
second jump the sample undergoes a transition into a FM state with the
magnetoresistance showing a saturated behavior at higher fields. For $%
x=0.125 $ these jumps are not clearly separated, a fact that has also been
observed in the magnetization curves. In contradiction to the $x=0.1$
compound, the magnetoresistance of the $x=0.125$ sample decreases with
increasing field below the field induced transition to the charge ordered FM
state. Finally for the $x=0.15$ sample (not shown) only negative CMR effects
have been observed in agreement with previous published results \cite
{Urushibara}. It is remarkable that in the temperature range between T$_{CA}$
and T$_{C}$ the magnetoresistance at a given field changes sign and becomes
positive.

To show these large positive MR effects, the temperature dependence of the
magnetoresistance for various fields and doping levels is plotted in Fig 3.
For $x=0.1$ two pronounced peaks yielding an increase in resistivity up to
+400\% when the magnetic field is raised to 5 T are clearly visible. These
peaks corresponds to the transitions into the O%
\'{}%
\'{}%
- and FM-state respectively. Around T$_{CA}$ the typical negative MR effect
can be seen. The same features are also present for $x=0.125$, though the
two positive enhachments at T$_{\text{O%
\'{}%
O%
\'{}%
\'{}%
}}$ and T$_{C}$ obviously merge resulting into a single peak. The positive
MR is maximal for $x=0.1$, becomes significant smaller for $x=0.125$ and
finally dissappears for $x\geqslant 0.15$. Summarizing the magnetoresistance
measurements indicate a large positive MR at the transitions O%
\'{}%
/CA$\rightarrow $O%
\'{}%
\'{}%
/CA and O%
\'{}%
\'{}%
/CA$\rightarrow $O%
\'{}%
\'{}%
/FM, while negative MR effects appear close to the O%
\'{}%
/PM$\rightarrow $O%
\'{}%
/CA and O/PM$\rightarrow $O/FM phase boundaries (see Fig. 1). The negative
MR obviously result from the onset of spin order below T$_{CA}$ (i.e. T$_{C}$
for $x>0.15$) and can be explained within a double-exchange picture as
proposed by \cite{Zener}. The electronic properties in the O%
\'{}%
\'{}%
-phase are more complicated and can not be explained taking only
double-exchange interactions in account.

Finally in Fig. 4 we show a H-T-phase diagram for La$_{0.9}$Sr$_{0.1}$MnO$%
_{3}$. We choose this concentration as for this sample the sequence of two
strongly coupled phase transitions (O%
\'{}%
$\rightarrow $O%
\'{}%
\'{}%
, CA$\rightarrow $FM) can be easily documented (see e.g. Figs. 2 and 3). The
insulating region of positive MR is enbedded in insulating phases which
reveal negative MR effects. Both transition temperatures (T$_{\text{O%
\'{}%
O%
\'{}%
\'{}%
}}$, T$_{C}$) are shifted to higher temperatures as the external field is
increased. At the two closely related phase boundaries T$_{\text{O%
\'{}%
O%
\'{}%
\'{}%
}}$ and T$_{C}$ three degrees of freedom, namely the charges (Mn$^{3+}$/Mn$%
^{4+}$), the spins and the orbitals undergo an ordering process. The
structural phase transition T$_{\text{O%
\'{}%
O%
\'{}%
\'{}%
}}$ indicates charge ordering \cite{Yam97} and most probably also orbital
ordering.

We start our discussion summarizing the most important experimental results
presented here. It is obvious that the transition from the canted-AFM and
JT-distorted O%
\'{}%
-phase to the pseudocubic charge-ordered O%
\'{}%
\'{}%
-phase is intimately coupled to a transition into an insulating FM state.
The positive CMR effect observed in the vicinity of the commensurate doping
value $x=0.125$ counts for a very different picture as it can be given by
the interplay of JT-distortion and DE-interactions alone.

It is well know that the double degeneracy of the $e_{g}$ levels of $3d$
ions in an octahedral enviroment is lifted by an JT-distortion of the
lattice \cite{Jahn}, accompanied by an ordering of the orbitals in real
space. The possibility of orbital ordering in transition-metal ions due to
exchange interactions different than that resulting from JT-distortions was
first pointed out by Roth \cite{Roth} and has been extensively studied by
Kugel and Khomskii \cite{Kugel}. It has been shown that two transitions take
place, one into an orbitally ordered state and a second into a spin-ordered
state, both driven by SE-mechanisms. This SE differs from the ordinary one
due to the fact that each electron has four degrees of freedom, two orbital
states ($d_{z^{2}}$, $d_{x^{2}-y^{2}}$) and two spin states (spin-up,
spin-down). The presence of intra-atomic exchange in case of orbital
degeneracy produce ferromagnetism below the orbital-ordering phase
transition \cite{Inagaki}. It is important to notice that above the
orbital-ordering transition temperature the effective spin-spin interaction
is AFM, which is modified by the appearance of the orbital ordered state and
finally goes over into a FM coupling. A modern view of this problem has been
recently considered by Held and Vollhardt \cite{Held}.

For undoped LaMnO$_{3}$ the magnetic properties can be well explained taking
only the predominant JT-distortion of the MnO$_{6}$ octaherda into account.
The double degeneracy of the $e_{g}$ orbitals is lifted by a long range
cooperative Jahn-Teller distortion resulting in an ordering of the $%
d_{z^{2}} $ orbitals as has been argued by Solovey $et$ $al$. \cite{Solovyev}
and has recently been confirmed experimentaly by Murakami $et$ $al.$ \cite
{Murakami}. As a result of this JT driven orbital ordering the A-type AFM
state is established below T$_{N}$.

Upon doping with holes the long-range JT-distortion become suppressed and
concomitantly double-exchange interactions have to be taken into acount. At
room temperature the lattice is distorted as in LaMnO$_{3}$ (O%
\'{}%
-phase) due to the JT effect removing the double degeneracy of the $e_{g}$
levels. Since mobile holes are present, the double exchange mechanism plays
a fundamental role. The magnetic structure observed in undoped LaMnO$_{3}$
is modified by the appearance of three-dimensional ferromagnetic
DE-interactions, competing with the JT driven A-type AFM yielding a canted
AFM state (or mixed phase). Thus canted antiferromagnetism is established
below T$_{CA}$=150 K, accompanied by a drop in the resistivity, since a gain
in kinetic energy of the carriers due to DE-interactions can be achieved.
Further lowering of the temperature favours SE-interactions between Mn$^{3+}$
ions in a manner discussed above, yielding the ordering of orbitals and
subsequently the evolution of ferromagnetism. At T$_{\text{O%
\'{}%
O%
\'{}%
\'{}%
}}$ the SE-interactions become dominant, suppress the JT-distortion and
enforces the structural transition into the pseudocubic O%
\'{}%
\'{}%
-phase, followed by the onset of ferromagnetism at T$_{C}$. At the same time
an insulating behavior appears due to charge and orbital ordering, which
decreases the hole mobility and explains the positive jumps in the MR
curves. The transition to the orbital ordered FM state, is stabilized by the
application of an external magnetic field as can be seen in Fig. 1 and Fig.
2. We remind that by application of an external magnetic field jumps in the
magnetization curves has been observed, leading the sample to higher
magnetization states \cite{Paras}. This field induced transitions can not be
ascribed to magnetocrystalline anisotropy since no dependence was found for
different orientations of the crystal axes in respect to the applied
external field. The possibility of magnetic field induced transitions due to
the interplay of the JT-effect and SE-interactions has been first pointed
out by Kugel and Khomskii \cite{Kugel2}. It was shown that when orbital
ordering is enforced by the SE mechanism such transitions are possible. On
the other hand if orbital ordering is established due to electron-lattice
interactions (JT-effects) as it is believed to be in LaMnO$_{3}$, then such
transitions may not appear.

At the moment only predictions about how the orbitals are ordered in the low
temperature state can be made, but an alternation of occupied $d_{z^{2}}$
and $d_{x^{2}-y^{2}}$ orbitals on neighbouring Mn$^{3+}$-sites appear most
probably to us (Fig. 5). This would cause a displacement of the O$^{2-}$
ions due to secondary electrostatic interactions \cite{Kugel}, yielding
alterning long and short Mn-O bonds. Together with the real space ordering
of the Mn$^{3+}$/Mn$^{4+}$ ions this would result in additional
superstructure reflections in the diffraction patterns. Finally for doping
levels higher than $x\geqslant 0.175$ the double exchange mechanism becomes
dominant and a metallic ferromagnetic state is established. The orbital
liquid picture recently proposed by Ishihara $et$ $al$. \cite{Ishihara} may
be appropriate to describe the physics in the metallic phase.

The authors would like to thank D. I. Khomskii for helpfull discussions.
This work has in part be supported by the BMBF under the contract number
13N6917.



FIGURE\ CAPTIONS:

Fig. 1: Phase diagram of La$_{1-x}$Sr$_{x}$MnO$_{3}$ at low doping
concentrations. The shaded area represents an orbitally and charge ordered
insulating ferromagnet.

Fig. 2: Isothermal magnetoresistance curves for the $x=0.1$ and $x=0.125$
samples.

Fig. 3: Temperature dependence of the MR for $x=0.1$ and $x=0.125$ at
various applied magnetic fields.

Fig. 4: H-T-phase diagram for $x=0.1$. The shaded area corresponds to the
region of positive MR. A strong negative MR appears close to the OO%
\'{}%
-phase boundary.

Fig. 5: Orbital ordering in La$_{0.875}$Sr$_{0.125}$MnO$_{3}$ which is in
accord with the charge order proposed by Yamada $et$ $al$.\cite{Yam97}.


\begin{references}
\bibitem{Chahara}  K. Chahara {\em et al.}, 
Appl. Phys. Lett. {\bf 63}, 1990 (1993); R. von Helmot {\em et al.}, 
Phys. Rev. Lett. {\bf 71}, 2331 (1993); S. Jin {\em et al.}, 
Science {\bf 264}, 413 (1994).

\bibitem{Zener}  C. Zener, Phys. Rev. {\bf 82}, 403 (1951).

\bibitem{deGennes}  P.G. de Gennes, Phys. Rev. {\bf 118}, 141 (1960).

\bibitem{Furukawa}  N. Furukawa, J. Phys. Soc. Jpn. {\bf 63}, 3214 (1994).

\bibitem{Millis}  A. J. Millis, P.B. Littlewood, and B.I. Shraiman, Phys.
Rev. Lett. {\bf 74}, 5144 (1995)

\bibitem{Paras}  M. Paraskevopoulos {\em et al.}, submitted to PRL.

\bibitem{Koshibae}  W. Koshibae {\em et al.}, J. Phys. Soc. Jpn. 66, 2985
(1997); T. Mizokawa and A. Fujimori, Phys. Rev. B 56, R493 (1997).

\bibitem{Ahn}  K. H. Ahn and A. J. Millis, Phys. Rev. B {\bf 58}, 3697
(1998).

\bibitem{Maezono}  R. Maezono, S. Ishihara and N. Nagaosa, Phys. Rev. B 58,
11583 (1998).

\bibitem{Mukhin}  A.A. Mukhin {\em et al.}, 
JETP Letters {\bf 69}, 356 (1998).

\bibitem{Yam97}  Y. Yamada {\em et al.}, 
Phys. Rev. Lett. {\bf 77}, 904 (1996).

\bibitem{Urushibara}  A. Urushibara {\em et al.}, 
Phys. Rev. B {\bf 51}, 14103 (1995).

\bibitem{Jahn}  H. A. Jahn and E. Teller, Proc. R. Soc. A{\bf \ 161, }220
(1937); H. A. Jahn, Proc. R. Soc. A {\bf 164}, 117 (1938).

\bibitem{Roth}  L. M. Roth, Phys. Rev. {\bf 149}, 306 (1966); J. Appl. Phys. 
{\bf 38}, 1065 (1967).

\bibitem{Kugel}  K. I. Kugel and D. I. Khomskii, JETP Letters {\bf 15}, 446
(1972); JETP {\bf 37}, 725 (1973).

\bibitem{Inagaki}  S. Inagaki, J. Phys. Soc. Jap. {\bf 39}, 596 (1975).

\bibitem{Held}  K. Held and D. Vollhardt, Eur. Phys. J. B 5, 473 (1998).

\bibitem{Solovyev}  I. Solovyev, N. Hamada, and K. Terakura, Phys. Rev.
Lett. {\bf 76}, 4825 (1996).

\bibitem{Murakami}  Y. Murakami {\em et al.}, 
Phys. Rev. Lett. {\bf 81}, 582 (1998).

\bibitem{Kugel2}  K. I. Kugel and D. I. Khomskii, JETP Letters {\bf 23}, 237
(1976).

\bibitem{Ishihara}  S. Ishihara, M. Yamanaka and N. Nagaosa, Phys. Rev. B
55, 4206 (1997).

\end{references}
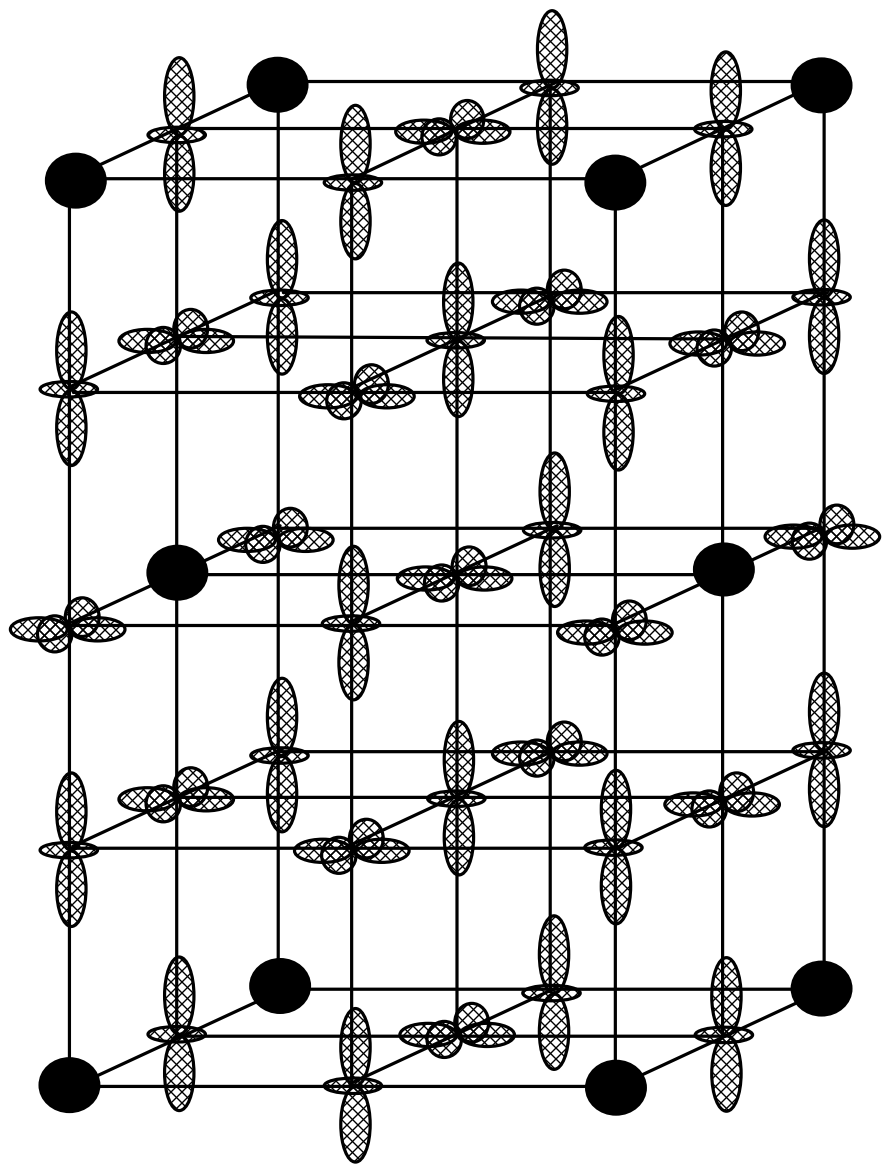
\end{document}